\documentstyle[prc,preprint,tighten,aps]{revtex}
\begin {document}
\draft
\preprint{\vbox{\noindent 
 \hfill LA-UR-97-4992\\
 \null\hfill nucl-th/9806074}}
\title{
R-Matrix and K-matrix Analysis of Elastic $\alpha$-$\alpha$ Scattering
}
\author{J. Humblet$^{a,b}$, A. Cs{\'o}t{\'o}$^{c,d,e}$ and K. Langanke$^{d,e}$
}
\address{
$^a$W.~K. Kellogg Radiation Laboratory, California 
Institute of Technology, Pasadena, CA 91125, USA\\
$^b$ Institute of Physics of the University, Sart Tilman B5, 4000 Liege 1,
Belgium \\
$^c$Theoretical Division, Los Alamos National Laboratory, Los Alamos, NM
87544, USA 
\\ 
$^d$ Institute for Physics and Astronomy, University of Aarhus, Aarhus, Denmark\\
$^e$ Theoretical Astrophysics Center, University of Aarhus, Aarhus,
Denmark  
}
\date{June 24, 1998}

\maketitle

\begin{abstract}
\noindent
The R- and K-matrix parametrizations are analyzed and compared for the
elastic $\alpha$-$\alpha$ scattering at center-of-mass energies below 40 MeV.
The two parametrizations differ in their definitions of the resonance
energy which can lead to quite different results. The physical values of
the best-fit parameters are compared with those computed for a potential
model. The existence of a broad resonance near 9 MeV is not supported by
the data or by the potential model. We discuss the positive and
negative aspects for both parametrizations.
\end{abstract}

\narrowtext

\section{Introduction}

The astrophysical S-factor for the reaction $^{12}$C($\alpha,\gamma$)$^{16}$O
at the astrophysically most effective 
energy of E=0.3 MeV has been obtained from the extrapolation
of a parametrized cross section by fitting data in the 
center of mass energy range between
1 and 3 MeV. Although Azuma {\it et al.} have recently found R-matrix
and K-matrix parametrizations which give nearly the same results
\cite{Azuma}, this
agreement has not always been observed and fits with quite  distinct 
differences between K- and R-matrix 
have been found 
\cite{Barker01,Barker02,Barker03,Ji,Humblet1,Filippone,Humblet2,Brune}. 
This situation has motivated
us to make detailed tests of the two parametrizations and to compare the
physical values from the two parametrizations. To this end, one must turn to
a much simpler problem than the simultaneous parametrization of the three
data sets for $^{12}$C$(\alpha,\gamma$)$^{16}$O, 
$^{12}$C($\alpha,\alpha$)$^{12}$C and $^{16}$N $\beta$-decay. Following Barker
\cite{Barker04} we consider the s-wave $\alpha+\alpha$ elastic scattering.
This problem also has the advantage that, besides accurate data,  an
excellent potential model description \cite{Friedrich} 
is available which we can use as a
benchmark. Below the energy of the first reaction threshold,
$^{7}$Li+p at 17.3 MeV, the 43 data we use 
\footnote{
Kindly communicated by F.C. Barker.
} 
are the same as
in reference \cite{Barker04}. But, when it is useful to consider higher
energies, we supplement them by the real parts of 5 complex phase
shifts obtained by Darriulat {\it et al.}  \cite{Darriulat}, as has been
done in \cite{Friedrich}, in the energy range 26-40 MeV. In the present
paper, all energies refer to the center-of-mass system.

The phase shifts corresponding to the potential model are obtained from
the radial wave function $u(r,E)$ which solves the radial Schr\"odinger
equation
\begin{equation}
\left[ \frac{d^2}{dr^2} - \frac{2M}{\hbar^2} \left( V_N+V_C-E \right)
\right] u(r,E)=0
\end{equation}
subject to the boundary condition
\begin{equation}
u(r=0,E)=0 \ \ .
\end{equation}
Here the nuclear and Coulomb potentials have the form \cite{Friedrich}
\begin{equation}
V_N(r) = V_0 \exp \left( -b r^2 \right) \ \ ;
\end{equation}
\begin{equation}
V_C(r)=4 e^2 {\rm erf} \left( \beta r \right) / r \ \ ,
\end{equation}
respectively.
The best fit to the data is obtained with $b=0.212$ fm$^{-2}$ and
$\beta=0.75$ fm$^{-1}$. The depth $V_0$ will be chosen in such a way
that the R-matrix and K-matrix parametrizations have a pole at the
energy of the $^8$Be ground state, i.e. at the energy $E_g=92.08$ keV
\cite{Barker04}.

\section{R-matrix fits}

In terms of the R-matrix \cite{Lane}, the s-wave phase shifts are
\begin{equation}
\delta(E) = - \phi(a,E) + {\rm arctan} \left(
\frac{P(a,E)}{R^{-1}+B-S(a,E)} \right) \ \ ,
\end{equation}
where $a$ is the channel radius, $B$ the real boundary condition
constant, and
\begin{equation}
\phi(a,E) = {\rm arctan} (\frac{F_0}{G_0} ) \ \ ,
\end{equation}
\begin{equation}
P(a,E)=\frac{\rho}{(F_0^2+G_0^2)} \ \ ,
\end{equation}
\begin{equation}
S(a,E)={P(a,E)} \ {(F_0 F'_0 + G_0 G'_0)}.
\end{equation}
$-\phi$, $P$, and $S$ 
are the hardsphere phase shift, the penetration factor and the shift
function, respectively. The radial Coulomb wave functions $F_0,G_0$
depend on
\begin{equation}
\rho={kr}=0.309428 r E^{1/2} ; \eta=0.891132 E^{-1/2} \ \ ,
\end{equation}
where $r$ is in fm and E is in MeV, while in Eq. (8) the primes stand
for $d/d\rho$. For the sake of comparison we adopt the radius parameter
$a=6$ fm from Ref. \cite{Barker04}. (We note that  this value appears to be
somewhat large. For example, Fowler {\it et al.} used $a=4$ fm
\cite{Fowler} and the conventional value is $a=1.44
(A_1^{1/3}+A_2^{1/3})=4.57$ fm. On the other hand, with $a=6$ fm, the
phase shifts have not quite reached their asymptotic values, as we will see
below.)

In terms of the radial factor $u(r,E)$, the R-matrix is
\begin{equation}
R=\frac{u(a,E)}{\left[ a (du/dr)_{r=a} - B u(a,E) \right]} \ \ .
\end{equation}
Let us call  $E_1,E_2,...$ the eigenenergies satisfying the boundary condition
\begin{equation}
a(du/dr)_{r=a} - B u(a,E) =0 \ \ .
\end{equation}
Together with (1) and (2) this defines a classical Sturm-Liouville problem
\cite{Morse} whose eigenvalues are all real. The R-matrix has been
defined in such a way that it has poles at the $B$-dependent
eigenenergies $E_i$. Its expansion then reads
\begin{equation}
R=\sum_{i=1}^{\infty} \frac{\gamma_i^2}{(E_i-E)} \ \ ,
\end{equation}
where the $\gamma_i^2$ are the formal 
\cite{Lane}
reduced widths, in
terms of which the formal widths are
\begin{equation}
\Gamma_i = 2 \gamma_i^2 P(a,E) \ \ .
\end{equation}

Note that the $E_i$ are not resonance energies. However, in R-matrix
theory a resonance is associated with each eigenenergy. Let us first
consider the case of a one-pole approximation, $R=\gamma_i^2 / (E_i-E)$.
We then have
\begin{equation}
\delta + \phi = {\rm arctan} \left( \frac{\Gamma_i/2}{E_i +
\Delta_i(a,E)-E} \right) \ \ ,
\end{equation}
where the energy shift is
\begin{equation}
\Delta_i (a,E) = \gamma_i^2 \left[ B - S(a,E) \right] \ \ .
\end{equation}
The resonance energy $E_{i,r}$ is defined as the shifted $E_i$
satisfying the equation \cite{Lane}
\begin{equation}
E_i + \Delta_i (a, E_{i,r}) - E_{i,r} =0 \ \ .
\end{equation}
Assuming that in the neighborhood of $E_{i,r}$ a linear approximation of
$\Delta_i(a,E)$ is satisfactory, the so-called observed \cite{Lane}
reduced width and observed width are
\begin{equation}
(\gamma_i^0)^2 = \frac{\gamma_i^2 }{[1+ \gamma_i^2 (dS/dE)_{E=E_{i,r}}]}
\ \ ,
\end{equation}
\begin{equation}
\Gamma_i^0 = 2 (\gamma_i^0)^2 P(a,E) \ \ ,
\end{equation}
and we have 
\begin{equation}
\delta + \phi = {\rm arctan} \frac{\Gamma_i^0 /2}{E_{i,r}-E} \ \ .
\end{equation}

For the general theory following from Eqs. (5) and (10) we have
\begin{equation}
\delta(E) + \phi(a,E) = {\rm arctan} \left[  P(a,E) Q(a,E)  \right]
\end{equation}
with
\begin{equation}
Q(a,E)=\frac{1}{(R^{-1}+B-S)}=\frac{u(a,E)}{a(du/dr)_{r=a}-S(a,E)
u(a,E)} \ \ .
\end{equation}
The function $Q(a,E)$ is neither an R-function in the sense of 
Wigner\footnote{
See Ref. \cite{Lane} p. 277 and the references there to
E.P. Wigner's original papers on the mathematical R-functions.
}
nor a meromorphic function of $E$. As a generalization of
Eq. (16), the resonances are defined as the real energies at which
\begin{equation}
R^{-1} +B-S=0 \ \ ,
\end{equation}
and we will call them $E_j^{(j)}, j=1,2,...$. They are pole energies of
$Q(a,E)$ and
\begin{equation}
\delta(E_j^{(j)}) + \phi(a,E_j^{(j)}) = 90^\circ ({\rm mod} \ \ 180^\circ)
\ \ .
\end{equation}
The pole term of $Q$ corresponding to $E_i^{(i)}$ is
\begin{equation}
\frac{(\gamma_i^0)^2}{(E_i^{(i)} -E)}
\end{equation}
with
\begin{equation}
(\gamma_i^0)^2=- \left[ (\frac{d}{dE} Q^{-1})_{E=E_i^{(i)}} \right]^{-1}
\ \ .
\end{equation}
The $E_{i,r}$ and $(\gamma_i^0)^2$ defined by Eqs. (16) and (17) are
one-level approximations of the generalized quantities $E_i^{(i)}$ and
$(\gamma_i^0)^2$ defined by Eqs. (22) and (25). When a good
potential model is available, $Q$ is given by Eq. (21). However,
generally this is not the case. Provided an R-matrix fit has been
obtained to the data, one can obtain the best values for $E_i^{(i)}$ and
$(\gamma_i^0)^2$ by substituting the parametrized R-function into Eqs.
(20) to (25).

When fitting data it is often convenient to choose the constant $B$ so
that the quantity $B-S(a,E)$ vanishes in Eq. (5) at one of the
energies $E_i^{(i)}$, say at $E_k^{(k)}$. When $B=S(a,E_k^{(k)})$, we
will call the eigenenergies which satisfy Eq. (11) by $E_1^{(k)},
E_2^{(k)},...$.

Turning to our example case, elastic $\alpha+\alpha$ scattering, we
choose $E_1^{(1)}=E_g$ and $B=S(6 \ \ {\rm fm}, E_g)$. Then the boundary condition
(11) is satisfied at $E=E_1^{(1)}$, if the potential depth is
$V_0=-119.216148$ MeV. Since $\phi(a=6 \ \ {\rm fm}, E_g)$ is very small (equal to
0.00029$^\circ$), we have to a very good approximation
\begin{equation}
\delta(E_g) =90^\circ \ \ ; (d \delta / dE)_{E=E_g} > 0
\end{equation}
and $E_g$ practically coincides with a resonance energy as defined by
the conventional definition.

In Fig. 1 the phase shifts obtained with the potential model are
compared with the data.
How do the poles in the best-fit
R-matrix compare with those of the potential model? To answer this
question we have computed the energies $E_i^{(j)}$, the reduced and the
formal widths for $i=1-3, j=1-3$ and the observed $\Gamma^0_i$ widths
for $i=1-3$. The results are in
Table 1, while we have summarized the best-fit R-matrix parametrizations
to the data below 18 MeV in Table 2, assuming $B=S(a=6 \ {\rm fm}, E_j^{(j)})$
with $j=1-3$. No widths can be attributed to $E_1^{(2)}$ and
$E_1^{(3)}$, which are below threshold. But it is easily verified that,
using the one-pole approximation with Eq. (16), $E_1^{(2)} =-0.395$
MeV in Table 2 is shifted to $E_{1,r}=0.101$ MeV, close to the resonance energy,
while using Eq. (22), $E_1^{(2)}$ is shifted exactly to the resonance
energy $E_g$. Similar results hold for $E_1^{(3)}=-0.379$ MeV.
In table 2, the three $\chi^2$ are different. This might be surprising
at first glance, as a transformation of the R-matrix parameters from one
boundary condition to another  
should not change the quality of the fits. However, this is only true if
{\it all} R-matrix parameters are allowed to vary \cite{Barker15}, which is not the case
here.

We also used the results in Table 1 
to calculate the R-matrix phase
shifts with a 
3-pole approximation for the case $B=S(a=6 \ {\rm fm}, E_1^{(1)})$.
In Fig. 2, these calculated phase
shifts are compared with the exact potential model 
phase shifts. The agreement is
not good. Comparing Tables 1 and 2, one observes that the reduced widths
$\gamma_3^2$ of the third poles in Table 2 (2.112, 1.892, and 1.821 MeV)
are much larger than those in Table 1 (0.734, 0.769, and 0.770 MeV).
Clearly, in the best fits (in Table 2), the large reduced widths at
the third poles compensate the contributions from poles at much higher
energies. The convergence of the sum of poles calculated for the
potential model is very slow. Even adding the fourth and fifth poles (at
$E_4^{(1)}=56.0$ MeV and $E_5^{(1)}=89.2$ MeV with the reduced widths
0.688 and 0.660 MeV, respectively) is not sufficient for a good fit, as
seen in Fig. 2.

Like $Q(a,E)$, the observed resonance energies $E_j^{(j)}$ and the
corresponding observed reduced widths $(\gamma_j^0)^2$ are
$B$-independent, but they also depend strongly on the radius $a$. This is
illustrated in Table 3 and Fig. 3 for the potential model. In Fig. 3,
with different channel radii, we plotted the phase $\delta+\phi$
corresponding to the data and to the potential model. The best fit of
the potential model to the 48 data is obtained with $a=5.5$ fm and
$\chi^2=84.1$. For each radius the potential depth $V_0$ has been
modified to satisfy exactly $E_1^{(1)}=E_g$. In the range $a=7-8$ fm,
the asymptotic phase has been reached and the fit remains good ($\chi^2
\leq 100)$. The density of resonances increases strongly  with the
channel radius without greatly changing $\chi^2$.

The uncertainty regarding the ``proper'' choice of the channel radius is such that,
in practice, it is often chosen to give the best fit to the data.

\section{K-matrix fits}

For the potential model the conventional K-matrix reads
\begin{equation}
{\rm K}=- \left(\frac{uF'_0 -u'F_0}{uG'_0-u'G_0} \right)_{r=a}
\end{equation}
and the phase shifts are
\begin{equation}
\delta(E) = {\rm arctan} {\rm K} \ \ .
\end{equation}
With the channel radius $a=6$ fm and the potential depth 
\footnote{
The potential depths $V_0$ are not the same in the R-
and K-matrix parametrizations. However, because the energy $E_g$ is very
small we  have $|F_0| << |G_0|$ and $S(a,E_g) \approx (\rho G'_0
/ G_0)$, so that the two depths are nearly the same.
} 
$V_0=-119.217576$ MeV, the K-matrix has a pole at $e_1=E_g=92.08$ keV, while other
real poles are at $e_2=2.808641$ MeV and at $e_3=31.881031$ MeV. But K
cannot be parametrized as a sum of pole terms, because of the essential
singularities of the radial Coulomb wave functions $F_0,G_0$. The
K-matrix has an infinite number of complex poles converging to $E=0$.

To eliminate these singularities, a modified ${\cal K}$-matrix has been
defined in Ref. \cite{Humblet3}. It has no other singularities than
isolated poles. It is obtained by separating the threshold factors of
$F_0,G_0$ which depend only on $\eta$, and substituting into $G_0$ a
polynomial in $\eta^{-2}$ (i.e. in E) to the singular function
\begin{equation}
h(\eta) = \frac{1}{2} \psi  (1+i\eta) + \frac{1}{2} \psi(1-i\eta) - {\rm
ln} \eta  \ \ ,
\end{equation}
where $\psi$ is the digamma function. The function $h(\eta)$ has an
essential singularity at $\eta=\infty$, i.e. at $E=0$. At real energies,
over a finite range (say for $E$ between $r_1$ and $r_2$), the
polynomial is chosen to fit the function $h(\eta)$ for $E \ge 0$ and
the function
\begin{equation}
h^+ (\eta) = \psi(i \eta) + \frac{1}{2i \eta} - {\rm ln} (i \eta)
\end{equation}
for $E \le 0$. This is easily achieved, at any desired approximation,
using Chebyshev polynomials up to the appropriate degree in $\eta^{-2}$.
Let $n$ be the maximum degree chosen for these polynomials. For a given
$n$, one can then choose $r_1$ and $r_2$ so that the polynomial in
$\eta^{-2}$ is exactly equal to $h(\eta)$ at $E =e_1$ and $e_2$. With
$n=95$ and 
\begin{equation}
r_1=-4.874644 {\rm MeV}  \:; \ r_2=44.591714 {\rm MeV}
\end{equation}
the real poles $e_1,e_2$ are the same for K and ${\cal K}$, and the
$e_3$ pole is only very slightly shifted to $e_3=31.881027$ MeV.

The energy $e_1$ of the first pole satisfies the usual resonance
conditions
\begin{equation}
\delta(e_i) =90^\circ ({\rm mod} \ \  180^\circ) \:,\ (\frac{d \delta}{dE})_{E=e_i} > 0
\end{equation}
and the reduced width $g_1^2$ is positive (see Table 4). For the two
other poles, the energies $e_2$, and $e_3$ satisfy only the first condition.
Their reduced widths are negative. They are echo poles and contribute
only to the background of ${\cal K}$.

The ${\cal K}$-matrix corresponding to the potential model is the
meromorphic function
\begin{equation}
{\cal K}=- \left( \frac{u {\bar F'_0} - u' {\bar F_0}}{u {\bar G'_0} -u'
{\bar G_0}} \right)_{r=a} \ \ ,
\end{equation}
where ${\bar F_0}, {\bar G_0}$ are the modified Coulomb wave functions
\cite{Humblet3}, while for the phase shifts, we have
\begin{equation}
\delta(E) = {\rm arctan} (p^2 {\cal K} )
\end{equation}
with 
\footnote{
In order to have dimensionless ${\cal K}$ and reduced
widths with an energy dimension, we substitute $\eta^{-l-1/2}$ for
$k^{l+1/2}$ following the definition of $p$ by Eq. (4.7) in Ref.
\cite{Humblet3}.
}
\begin{equation}
p^2 = \frac{2 \pi}{(\exp \{2 \pi \eta \} -1 )}
\end{equation}

The very way we have defined the ${\cal K}$ matrix has several
consequences. Contrary to K, ${\cal K}$ can be expanded in a series of
pole terms
\begin{equation}
{\cal K} = \sum_{i=1}^{\infty} \frac{g_i^2}{(e_i-E)} \ \ .
\end{equation}
At positive real energies, the phase shifts defined by Eqs. (28) and
(34) are practically equal, while, at complex energies, ${\cal K}$ is
different from $K/p^2$. Like the polynomial in $\eta^{-2}$, ${\cal K}$
also depends on the choice of $n, r_1$, and $r_2$.

With the potential model, the three real poles of ${\cal K}$ below 40
MeV and their reduced widths are given in Table 4. With the
corresponding 3-pole approximation of ${\cal K}$, one does not obtain a
good fit of the phase shifts. A better approximation requires more real
and/or complex poles. Like the expansion of the R-matrix, the expansion
of ${\cal K}$ converges only very slowly, and the existence of complex
poles introduces further complications. Moreover, only the real poles
are uniquely defined, in the sense that the energy of the complex poles
and their residues depend on the choice of $n, r_1$ and $r_2$. In
particular, the density of complex poles near the real axis increases
when a larger  $n$ is chosen.

In the domain $-$4 MeV $\leq {\rm Re} E \leq$ 44 MeV, $-$ 5 MeV $\leq
{\rm Im} E \leq 5$ MeV with $n=45,70,95$ the number of pairs of complex
conjugate poles is 7, 22, and 37, respectively. Many of these poles have very
small complex residues and their contributions to the ${\cal K}$-matrix
are negligible.

In a ${\cal K}$-matrix fit to the data, when an additional background
term is needed, it cannot be a real pole in the energy range of the
data, since $\delta =90^\circ ({\rm mod} \  180^\circ)$ 
only at $e_1,e_2,e_3$. But a real
background pole at negative energy cannot be excluded a priori. In fact,
the energy dependence of the ${\cal K}$-matrix background plotted in
Fig. 2 of \cite{Barker04} shows a decreasing positive background with a
concave curvature. Since the energy dependence of an echo pole below
threshold has also a concave curvature at positive energies, introducing
such a pole into the parametrization seems to be the simplest way to get
a good fit below 18 MeV. With the $e_1$ and $e_2$ poles and a
3-parameter background composed of a constant and an echo pole at
$-$0.327 MeV, Barker \cite{Barker04} easily obtained a very good fit to
the data. We checked that the ${\cal K}$ matrix for the potential model
has no such pole at negative energies and we must conclude that the
three parameters of the background 
are only {\it ad hoc} parameters without physical meaning.

To further confirm this conjecture, we made another best fit to the data
up to 40 MeV using the three real poles and a pair of complex conjugate
poles. With $e_4=e_R+ie_I$ and $g_4^2 = \gamma_R + i \gamma_I$, the
background term has the form
\begin{equation}
\frac{g_4^2}{2(e_4-E)} + {\rm conj.} = \frac{\gamma_R (e_R-E) + \gamma_I
e_I}{(e_R-E)^2 + e_I^2} \ \ .
\end{equation}
A very good fit is obtained to the 48 data points up to 40 MeV
($\chi^2=27.0$).
The results are in Table 4 and Fig. 4. They confirm that in a ${\cal K}$
matrix fit to data with a minimal number of parameters, the background
parameters are unlikely to correspond to the  complex poles computed
for a phase-equivalent potential model.

\section{Conclusions}

The main advantage of an R-matrix parametrization is that $R$ can be
expanded in terms of real pole terms because $R$ is an $R$-function in
the sense of Wigner. The parameters introduced in an R-matrix fit are
the eigenenergies $E_i$ and the formal reduced widths $\gamma_i^2$. 
However,
this
is a parametrization of $\delta+\phi$ rather than of $\delta$. One sees
in Fig. 3 that at $E\geq 2$ MeV $\delta+\phi$ is steadily increasing
with $E$ and one can obtain the energies, at which
\begin{equation}
\delta (E) + \phi(a,E) = 90^\circ ({\rm mod} \ \ 180^\circ) \ \ ,
\end{equation}
defining the so-called observed R-matrix resonance energies. The
corresponding observed reduced widths $(\gamma_i^0)^2$ and the observed
widths $\Gamma_i^0$ are obtained from Eqs. (16) and (17), or, if
a 1-pole approximation
does not apply, from Eqs. (22) and (25).

Do all these so-called resonance energies $E_i^{(i)}$ really correspond
to physical resonances? How are they compared to the resonances defined
by the usual conditions
\begin{equation}
\delta (E) =90^\circ ({\rm mod} \ \ 180^\circ), \: \frac{d \delta}{dE} > 0
\ \ ?
\end{equation}
Let us first evaluate
the time delay in the
scattering process, which is $2 v^{-1} d\delta / dk$ \cite{Lane}, where $v=\hbar
k/M$ ($M$ being the reduced mass) and
\begin{equation}
\frac{d\delta}{dk} = \frac{d}{dk} \left[ {\rm arctan} (P(a,E)Q(a,E))
\right] - \frac{d}{dk} \phi (a,E) \ \ .
\end{equation}
With the potential model, 
at $E=E_1^{(1)}$, the two terms on the right hand
side of Eq. (40) are $61 \cdot 10^4$ and $97 \cdot 10^{-5}$ fm,
respectively, and the resonance conditions (39) are satisfied. At
$E=E_2^{(2)}$, the same terms are 3.7 and 6.6 fm, respectively. Thus,
$d\delta/dk$ is negative and this corresponds to a time advance,
excluding a physical resonance. A similar argument holds also at
$E=E_i^{(i)}$ when $i>2$, but in Ref. \cite{Barker04}, the third pole at
32.9 MeV was already considered as a background term.

Let us now see how the $E_1^{(2)}$ R-matrix pole appears as a complex
pole of the ${\cal S}$-matrix. Designating that pole by $p(E)$, we
obtained
\begin{equation}
p(E)=\frac{(-0.0535+ i 0.0339)}{(E_c-E)}
\end{equation}
where $E_c=3.34-i 16.59$ MeV is far off the real axis. With ${\cal S}
(3.34)=-0.921 + i 0.389$ and $p(3.34)=-0.00204-i 0.00322$, one finds
that $p(E)$ does not add a resonant contribution to ${\cal S}(E)$.

Under such conditions, we must conclude that, in the best fit,
$E_2^{(1)}$ and $E_3^{(1)}$ are both background poles, despite the fact
that $\gamma_2^2$ and $\gamma_3^2$ (like $\gamma_1^2$) are positive.
This has been confusing \cite{Barker04} and, like the strong dependence
of the $E_i^{(i)}$ on the channel radius, it is a weak point of the
R-matrix parametrization. This does not concern the good quality of fits
one can obtain with this method, but the physical interpretation of the
parameters of the pole terms.

In a ${\cal K}$ matrix parametrization, the approximate energies at
which there are real poles of ${\cal K}$ are directly suggested by the
data, since they are the energies at which $\delta=90^\circ ({\rm mod} 
\ \ 180^\circ)$.
This is the main advantage of a ${\cal K}$-matrix parametrization.
These energies are either resonances or echo pole energies according to
whether the phase shift data are increasing or decreasing at the 
energies concerned. 
In the present case, the echo poles at 2.8 MeV and 31.8 MeV
are only a part of the background terms of ${\cal K}$. At least one more
background term is needed to obtain a good fit to the data. It can be a
constant, a real echo pole below threshold or at higher energies than
the data, or a pair of conjugate complex energy poles. The parameters of
these poles, energies and residues, are no more than {\it ad hoc}
parameters. They are not expected to be poles of the modified 
${\cal K}$-matrix corresponding to a potential model. This is the weak point of a
${\cal K}$-matrix parametrization.

We can summarize as follows the lessons we have learned from the present
analysis of the $\alpha+\alpha$ scattering using the R-matrix and ${\cal
K}$-matrix methods for pratical fit procedures. In order to obtain a
good fit to data, one should use as flexible parametrizations as
possible, without worrying about the relation of certain parameters
(e.g. echo poles below threshold in the ${\cal K}$-matrix or R-matrix
eigenenergies and residues) to physical quantities (e.g. resonance
parameters). In certain cases such a connection cannot  be made and some
of
the fit parameters are purely {\it ad hoc}. Nevertheless, one of the
main purposes of R-matrix and ${\cal K}$-matrix fits is to give a
reasonable and physically motivated basis to extrapolate data. If both
methods are used carefully, they can be expected to give similarly good 
parametrizations, as in the case of $^{12}$C($\alpha,\gamma$)$^{16}$O.

\acknowledgements
The authors are grateful to C.A. Barnes, S.E. Koonin and R.D. McKeown
for useful discussions and the warm hospitality at the Kellogg Radiation
Laboratory of the California Institute of Technology. The authors also
thank C.A. Barnes and S.E. Koonin for a careful and critical reading of
the manuscript. 
Attila Cs{\'o}t{\'o}  and Karlheinz Langanke thank the Danish Research
Council for financial support.
The work of Attila Cs{\'o}t{\'o} was performed under the auspices of
the U.S. Department of Energy.

\narrowtext
\begin{figure}
\caption{
Comparison of the experimental phase shifts (points) with the potential
model (solid line). The dashed line corresponds to the best R-matrix
fits in Table 2. Note that the 3 best fits are indistinguishable within
the thickness of the line.
}
\label{fig1}
\end{figure}

\narrowtext
\begin{figure}
\caption{
Comparison of the potential model phase shifts (solid line) with the
computed 3-pole (long-dashed) and 5-pole (short-dashed) R-matrix approximations.
Compared with the data, the $\chi^2$ values are 93.9 (potential model),
9202 (3-pole) and 3897 (5-pole), respectively.
} 
\label{fig2}
\end{figure}

\narrowtext
\begin{figure}
\caption{ 
Values for $\delta+\phi$ for the data and the potential model to
illustrate the dependence of the R-matrix $E_i^{(i)}$ resonances on the
channel radius.
}
\label{fig3}
\end{figure}

\narrowtext
\begin{figure}
\caption{
${\cal K}$-matrix fits to the data. The solid curve (fitted to the data
up to 40 MeV) has been obtained with a pair of conjugate complex poles
as background, and the dashed line (fitted to the data up to 20 MeV)
with a background consisting of a constant and an echo pole below threshold. 
}
\label{fig4}
\end{figure}

\narrowtext
\begin{table}
\caption{With the potential-model, three different computed
parametrizations of the R-matrix are obtained with the
boundary-condition constants $B=S(6,E_i^{(i)})$ $(i=1,2,3)$,
respectively. All the parameters are in MeV.}
\begin{tabular}{lr@{}lr@{}lr@{}lr@{}lr@{}l}
i & \multicolumn{2}{c}{1} &\multicolumn{2}{c}{2} & 
  \multicolumn{2}{c}{3} & \multicolumn{2}{c}{4} & 
  \multicolumn{2}{c}{5} \\
\hline
$E_i^{(1)}$ & 0.&09208 & 9.&764 & 29.&274 & 56.&006 & 89.& 217 \\
$\gamma_i^{2}$ & 0.&228 & 0.&835 & 0.&734 & 0.&688 & 0.&660 \\
$\Gamma_i$ & 8.&15E--6 & 9.&207 & 14.&500 & 18.&947& 23.&022  \\
\hline
$E_i^{(2)}$ & --0.&343 & 8.&419 & 28.&173 &    &    &   &   \\
$\gamma_i^{2}$ & 0.&386 & 1.&004 & 0.&769 &   &    &   &    \\
$\Gamma_i$ & \multicolumn{2}{c}{---} & 10.&195 & 14.&903 & & & & \\
\hline
$E_i^{(3)}$ & --0.&352 & 8.&397 & 28.&156 &    &    &   &   \\
$\gamma_i^{2}$ & 0.&389 & 1.&006 & 0.&770 &   &    &   &    \\
$\Gamma_i$ & \multicolumn{2}{c}{---} & 10.&203 & 14.&904 & & & & \\
\hline
$(\gamma_i^0)^2$ & 0.&174 & 1.&000 & 0.&769 &   &    &   &    \\
$\Gamma_i^0$ & 6.&20E--6  & 10.&156 & 14.&901 & & & & \\
\end{tabular}
\end{table}

\narrowtext
\begin{table}
\caption{Best R-matrix fits below 18 MeV. The parameters in parentheses
denote fixed input values. The fixed parameters for the second and third
best fits are obtained using Eq.\ (21) with the parametrized R-matrix
from the first best fit. All the parameters are in MeV. The three
$\chi^2$ are not identical because not all parameters have been varied
\protect\cite{Barker15}.}
\begin{tabular}{lr@{}lr@{}lr@{}lr@{}l}
i & \multicolumn{2}{c}{1} &\multicolumn{2}{c}{2} & 
  \multicolumn{2}{c}{3} & \multicolumn{2}{c}{$\chi^2$} \\ 
\hline
$E_i^{(1)}$ & (0.&09208) & 9.&787 & 32.&954 & &  \\
$\gamma_i^{2}$ & (0.&199) & 0.&848 & 2.&112 & &  \\
$\Gamma_i$ & (7.&099E--6) & 9.&366 & 44.&359 & 18.&05    \\
\hline
$E_i^{(2)}$ & --0.&395 & (8.&424) & 30.&463 &   &   \\
$\gamma_i^{2}$ & 0.&343 & (1.&012) & 1.&892 &   &    \\
$\Gamma_i$ & \multicolumn{2}{c}{---} & (10.&281) & 38.&162 & 15.&86 \\
\hline
$E_i^{(3)}$ & --0.&379 & 8.&406 & (30.&032) &  &   \\
$\gamma_i^{2}$ & 0.&336 & 1.&010 & (1.&821) &  &    \\
$\Gamma_i$ & \multicolumn{2}{c}{---} & 10.&243 & (36.&459) & 15.&90 \\
\hline
$(\gamma_i^0)^2$ & (0.&156) & (1.&008) & (1.&820) &   &    \\
$\Gamma_i^0$ & (5.&57E--6)  & (10.&241) & (36.&441) & & \\
\end{tabular}
\end{table}

\narrowtext
\begin{table}
\caption{Nuclear potential depths, and second and third poles of
$Q(a,E)$, for different values of $a$ (in $fm$). The depths $V_0$ (in
MeV) are chosen so that $E^{(1)}_1=E_g$ for each radius $a$. $V_0$ and the
$E_i^{(i)}$ are in MeV. The $\chi^2$-values have been calculated for 48
data points.}
\begin{tabular}{r@{}lr@{}lr@{}lr@{}lc}
\multicolumn{2}{c}{$a$} & \multicolumn{2}{c}{$V_0$} &
  \multicolumn{2}{c}{$E_2^{(2)}$} & \multicolumn{2}{c}{$E_3^{(3)}$} & 
  $\chi^2$ \\ 
\hline
4.&0 & --122.&8034 & 28.&755 & 78.&050 & 1464\\
5.&0 & --119.&5106 & 14.&801 & 44.&871 & 143 \\
5.&5 & --119.&2750 & 11.&018 & 35.&186 & 84.1 \\
6.&0 & --119.&2161 &  8.&418 & 28.&155 & 93.9 \\
7.&0 & --119.&2003 &  5.&279 & 18.&980 & 100 \\
8.&0 & --119.&1999 &  3.&595 & 13.&516 & 99.7 \\
\end{tabular}
\end{table}

\mediumtext
\begin{table}
\caption{
${\protect\cal K}$-matrix
parameters computed from  the potential-model and two
best fits up to 18 and 40 MeV, respectively. Except for the
dimensionless constant in the first best fit, all the parameters are in MeV.}
\begin{tabular}{cr@{}lr@{}lr@{}lr@{}lr@{}lr@{}l}
 &\multicolumn{4}{c}{Potential model} & \multicolumn{4}{c}{Best fit} &
  \multicolumn{4}{c}{Best fit} \\ 
Range & \multicolumn{4}{c}{0.01--40 MeV} & \multicolumn{4}{c}{0.01--18 MeV} & 
  \multicolumn{4}{c}{0.01--40 MeV} \\
\hline
$j$ & \multicolumn{2}{c}{$e_j$} & \multicolumn{2}{c}{$g_j^2$} &
 \multicolumn{2}{c}{$e_j$} & \multicolumn{2}{c}{$g_j^2$} &
 \multicolumn{2}{c}{$e_j$} & \multicolumn{2}{c}{$g_j^2$} \\
1 & 0.&09208 & 46.&138 & (0.&09208) & (45.&726) & (0.&09208) & (45.&726) \\  
2 & 2.&809 & --10.&801 & 2.&819 & --10.&278 & 2.&809 & --10.&924 \\  
3 & 31.&881 & --6.&364 & --0.&324 & --39.&504 & 31.&865 & --5.&236 \\  
4 & & & & & \multicolumn{4}{c}{const.=--0.656}  & 
  $-1.026\pm$&$i0.199$ & $-31.082\pm$&$i307.53$ \\  
$\Gamma_1$ & \multicolumn{4}{c}{5.62E--6} & \multicolumn{4}{c}{(5.57E--6)}
& \multicolumn{4}{c}{(5.57E--6)} \\
$\chi^2$ & \multicolumn{4}{c}{92.5} & \multicolumn{4}{c}{18.2} & \multicolumn{4}{c}{27.0} \\
\end{tabular}
\end{table}


\begin{references}
\bibitem{Azuma} R.E.Azuma {\it et al.}, Phys. Rev. C50 (1994) 1194
\bibitem{Barker01} F.C. Barker, Austr. J. Phys. 24 (1971) 777
\bibitem{Barker02} F.C. Barker, Aust. J. Phys. 40 (1987) 25
\bibitem{Barker03} F.C. Barker and T. Kajino, Aust. J. Phys. 44 (1991)
369
\bibitem{Ji} X. Ji, B.W. Filippone, J. Humblet and S.E. Koonin, Phys.
Rev. C41 (1990) 1736
\bibitem{Humblet1}  J. Humblet, P. Dyer and B.A. Zimmerman, Nucl. Phys.
A271 (1976) 210
\bibitem{Filippone} B.W. Filippone, J. Humblet and K. Langanke, Phys.
Rev. C40 (1989) 515
\bibitem{Humblet2} J. Humblet, B.W. Filippone and S.E. Koonin, Phys.
Rev. C44 (1991) 2530 and C48 (1993) 2114.
\bibitem{Brune} C.R. Brune, Nucl. Phys. A596 (1996) 122
\bibitem{Barker04} F.C. Barker, Nucl. Phys. A575 (1994) 361
\bibitem{Friedrich} B. Buck, H. Friedrich and C.. Wheatley, Nucl. Phys.
A275 (1977) 246; H. Friedrich, Phys. Rep. 74 (1981) 209
\bibitem{Darriulat} P. Darriulat, G. Igo and H.G. Pugh, Phys. Rev. 137
(1965) B315
\bibitem{Lane} A.M. Lane and R.G. Thomas, Rev. Mod. Phys. 30 (1958) 257
\bibitem{Morse} P.H. Morse and H. Feshbach, {\it Methods of Theoretical
Physics}, Vol. 1, McGraw-Hill, New York 1953.
J.D. Pryce, {\it Numerical Solutions of Sturm-Liouville Problems}
Clarendon Press, Oxford, 1993.
\bibitem{Fowler} W.A. Fowler, G.R. Caughlan and B.A. Zimmerman, Ann.
Rev. Astron. Astrophys. 5 (1967) 525 and 13 (1975) 69; J. Humblet, W.A.
Fowler and B.A. Zimmerman, Astron. Astrophys. 177 (1987) 317 Table 1
\bibitem{Humblet3} J. Humblet, Phys. Rev. C42 (1990) 1582
\bibitem{Barker15} F.C. Barker, Aust. J. Phys. 25 (1972) 341
\end{references}
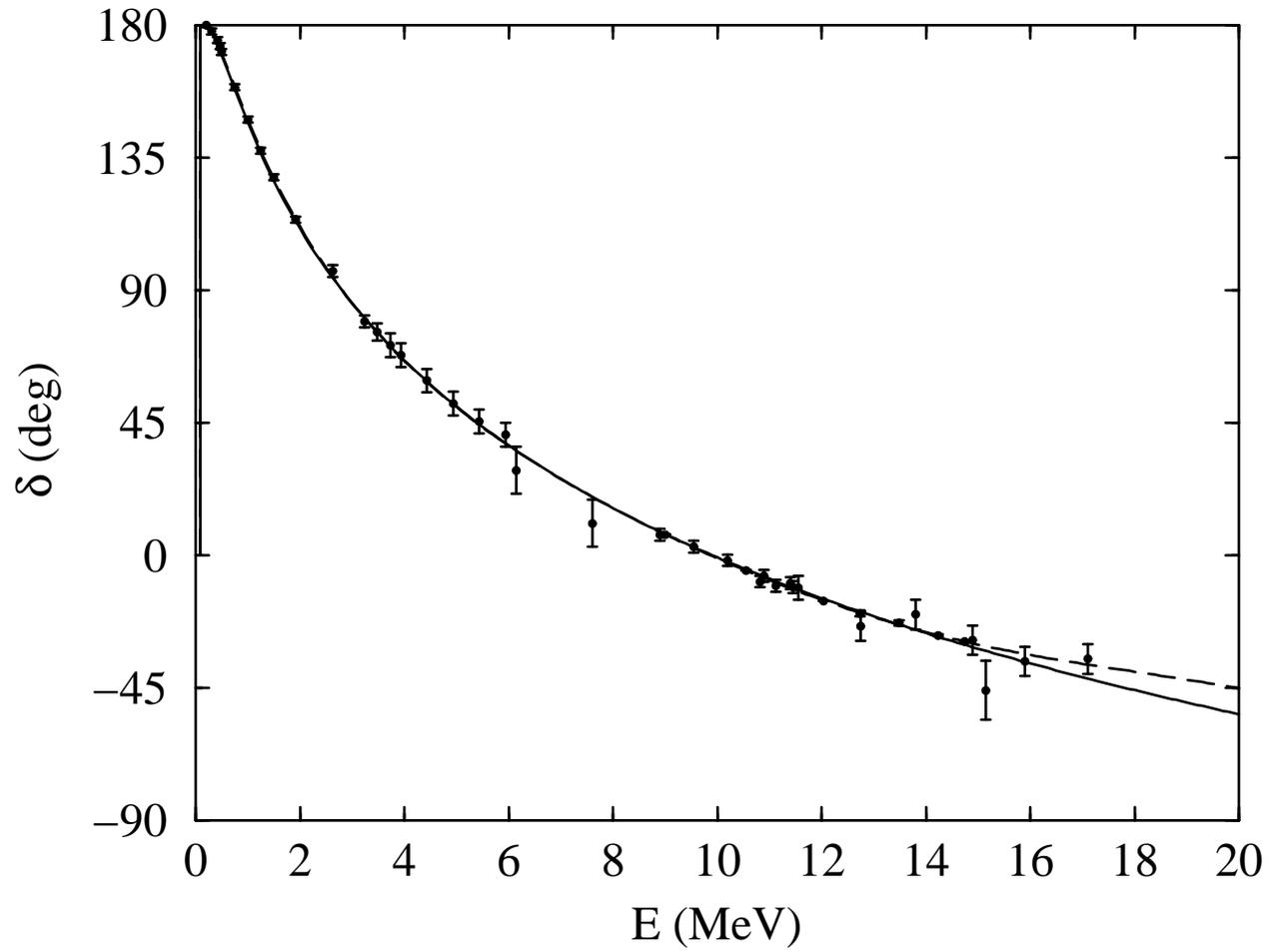
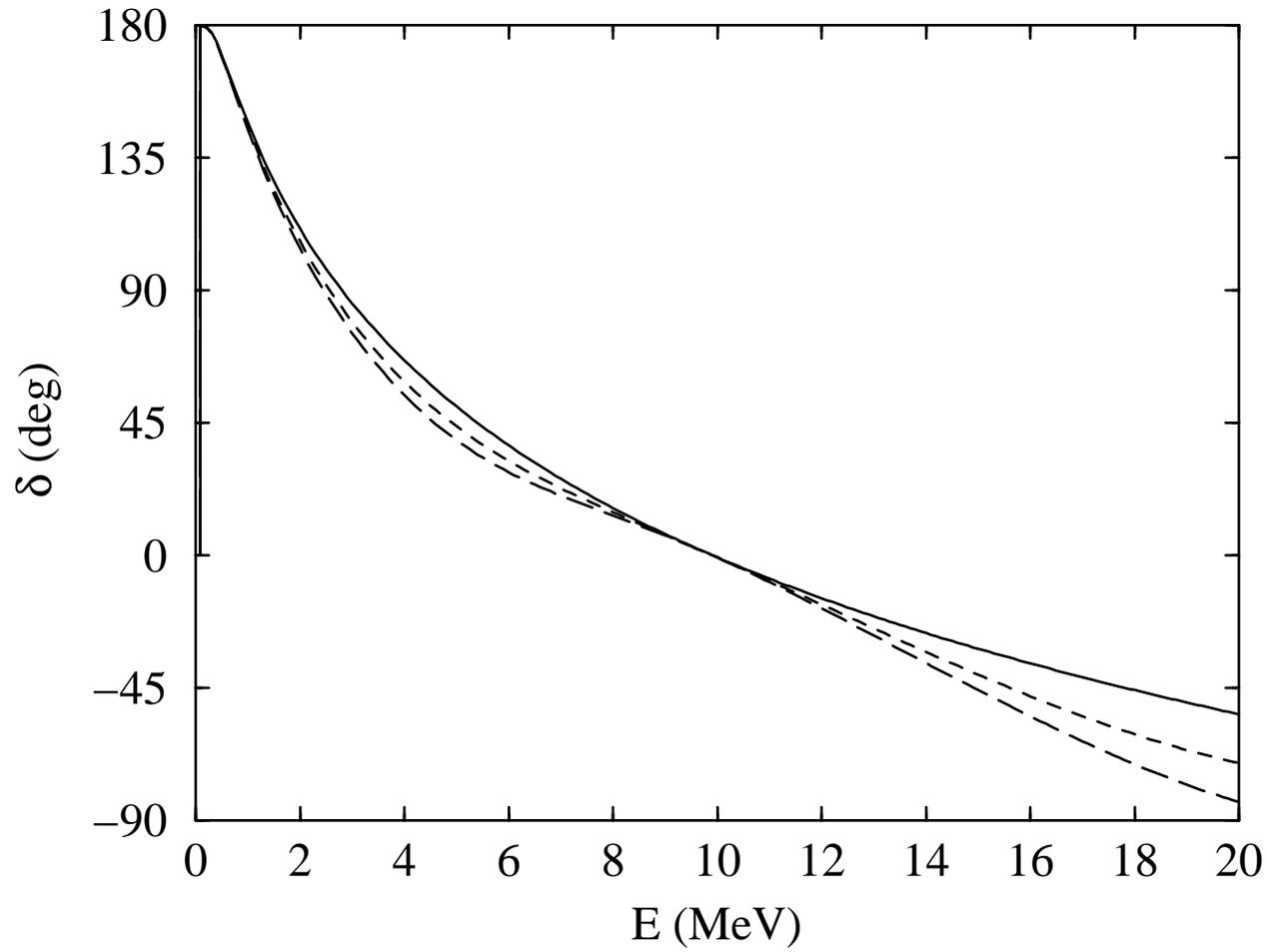
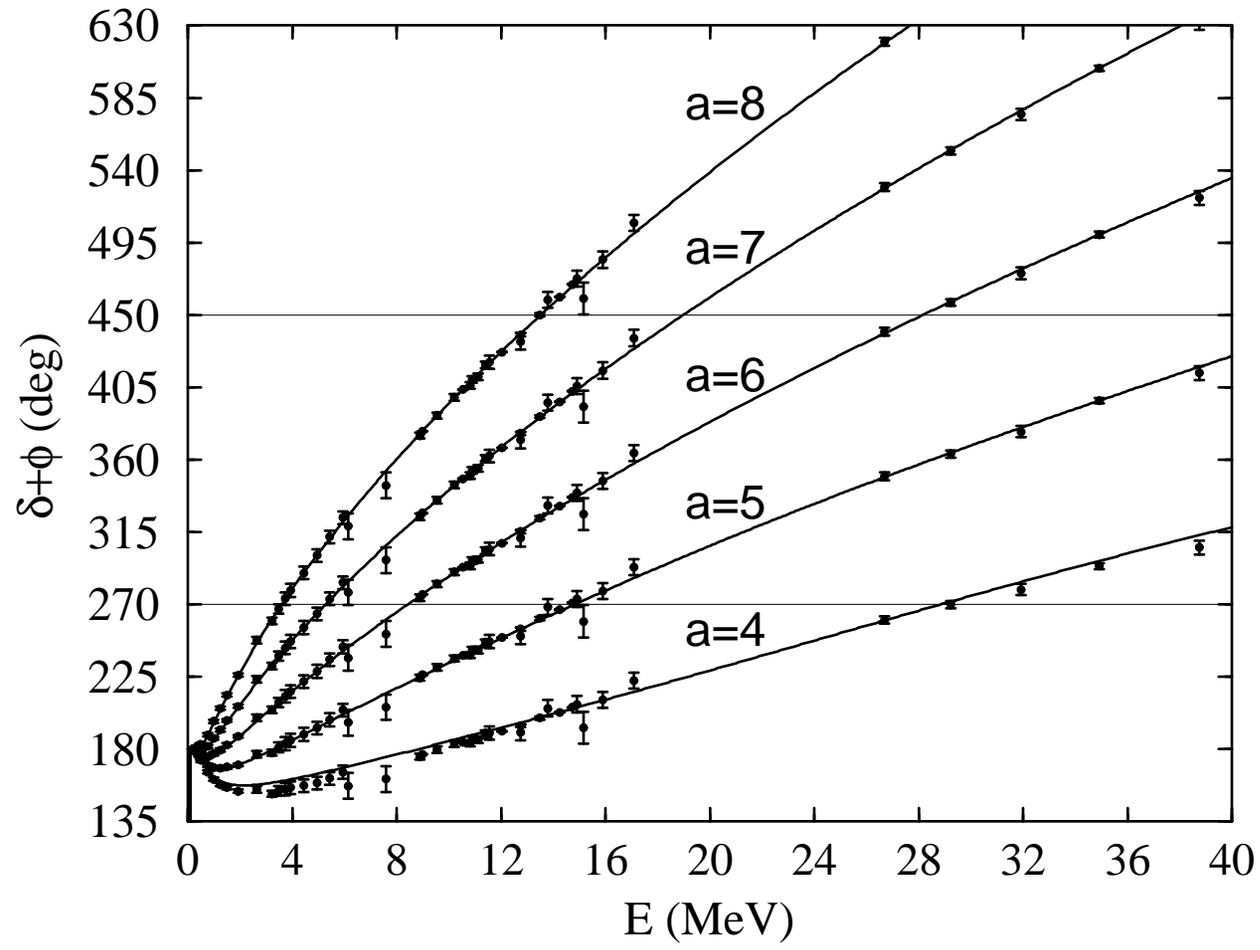
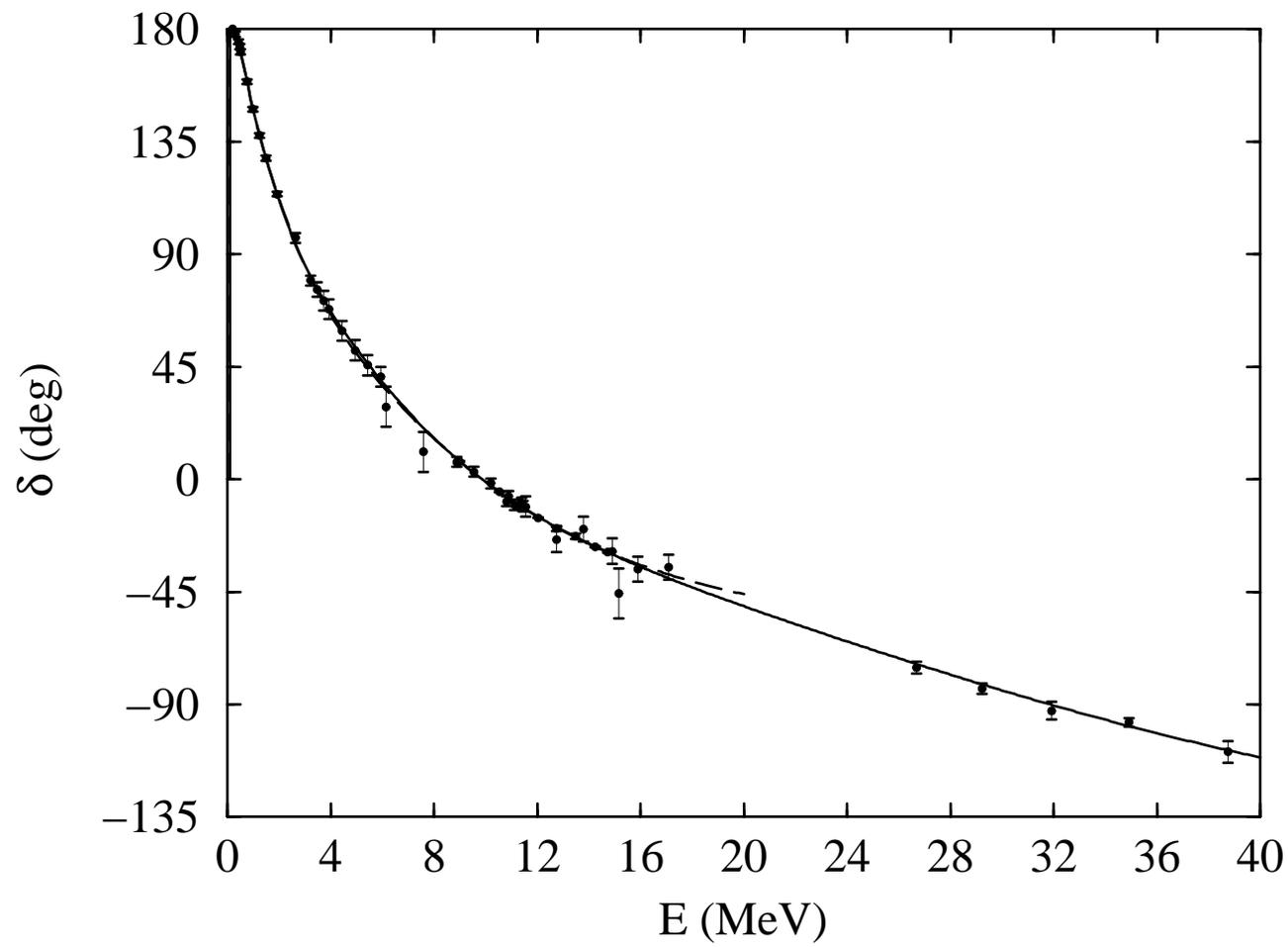
\end{document}